# A Computer Program for Objective Point Symmetry Classifications of Pseudosymmetric Electron Diffraction Spot Patterns

L. von Koch and P. Moeck

*Abstract*—A Python program for calculating the metrics necessary to perform information-theory based symmetry classifications and quantifications of transmission electron diffraction spot patterns is introduced. It is the first of its kind, in that it implements objectivity into crystallographic symmetry classifications and quantifications of approximate zone axis patterns from crystals. The equations by which the program operates as well as the required inputs are given. The results of the program's analysis of an experimental transmission electron diffraction spot pattern from a crystal with a pseudo-hexagonal lattice metric and a rectangular-centered Bravais lattice is used as an example. The program will eventually be appended to allow analysis of the other hierarchical translational pseudo-symmetry and Bravais lattice type combinations. Crystallographic $R_{sym}$ values of traditional classifications into projected point symmetry groups are provided alongside information-theoretic results of the new program's analysis for comparison purposes.

## I. Introduction and Background

Subjectivity permeates the traditional two-dimensional (2D) symmetry classifications in electron crystallography [1], organic chemistry [2], and computer science [3]. It was recently stated with respect to cryo-electron microscopy that, *"… as currently practiced, the procedure is not sufficiently standardized: a number of different variables (e.g. … threshold value for interpretation) can substantially impact the outcome. As a result, different expert practitioners can arrive at different resolution estimates for the same level of map details."* [4].

Although computer programs exist that provide symmetry deviation measures with respect to experimental electron diffraction spot patterns, e.g. CRISP/ELD with its subroutine Space Group Determinator [1], they contain subjective elements in that they are based on programmer-determined thresholds for symmetry interpretations and classifications.

Crystallographic symmetry classification and quantification methods [6-11] that employ Kanatani's geometric form of information theory [12] overcome this kind of subjectivity because they provide a workaround to the symmetry inclusion problem that has eluded computer scientists [3] for more than five decades [5]. Note that a lower symmetric geometric model always provides a better fit to experimental data than a higher symmetric model. This is because the former features more degrees of freedom as it is less restricted by symmetries. Optimizing the fit between geometric models of experimental data and the data itself in the traditional manner [1-3], can, therefore, not be a winning strategy for optimal model selection in the presence of symmetry inclusion relations and generalized noise.

Kanatani's geometric form of information theory allows for numerical quantifications of the evidence in favor and against each geometric model or group of models within a complete set of models. Relying only on numerical relations, i.e. inequalities, rather than on subjectively interpreted values of traditional residuals such as $R_{sym}$ values [1] leads to objectivity in the model selection process. The authors define objectivity as forming conclusions only on the basis of the information that is contained in the experimentally gathered data. Reasonable assumptions about the generalized noise in the data are required for an information-theoretic analysis; however, this does not take away from the objectivity of the model selection process.

Since translational pseudosymmetries are not rare in nature [13,14] and the Bravais lattice types are known to form a hierarchy with transitions from lower symmetric types to higher symmetric types that are marked by metric specializations [15], one should always use either a 2D point symmetry group [10,11] or plane symmetry group and projected 2D Laue class [8,9] classification as basis for the assignment of a 2D Bravais lattice type to an electron diffraction spot pattern or an atomic/molecular resolution direct space image, respectively. (Note that experimentally derived metric tensor components always feature error bars so that the derived lattice parameters never match the symmetry restriction of the given Bravais lattice type exactly.)

It has, for example, been inferred that the quaternary symmetry of an intrinsic membrane protein embedded in a lipid bilayer must be 3D point symmetry group *4* because there "seemed" to be a square 2D Bravais lattice type (within experimental errors) and plane symmetry group *p4gm* in an orthogonal projection in transmission electron microscopes (TEMs) [16-18]. No hard evidence for that assertion was, however, presented as their symmetry analyses contained elements of subjectivity.

An information theoretic analysis of some of the same direct-space phase-contrast TEM images that were used in [17] indicated that projected Laue class *2mm* and plane symmetry *p2gg* are more likely [9], identifying both the apparent four-fold rotation symmetry of the protein complex and the corresponding square Bravais lattice as strong pseudosymmetries [15]. An image/map data supported model mechanism for the opening and closing of this particular

The first author volunteers in Portland State University's Nano-Crystallography Group and is currently a senior of Westside Christian High School, Portland, Oregon, USA (email: lukasvonkoch@gmail.com). He will soon be a freshman at the University of Pennsylvania. The second author is with the Department of Physics, Portland, OR, USA (phone: 503-725-4227; fax: 503-725-2815; e-mail: pmoeck@pdx.edu).

membrane protein that is restricted to four-fold rotation symmetry (as the one in [18]) has, accordingly, at present less experimental support than an alternative mechanism that is restricted to incorporate two-fold rotation symmetry only.

In order to make further progress in the electron crystallography field in general and in specific cases such as the above mentioned example, one needs to embrace Kanatani's geometric form of information theory in the analysis of electron diffraction spot patterns [10,11]. To do this effectively one needs a computer program. This paper describes such a program that has been written by the first author and tested on experimental data that the second author provided. (The program can be obtained from the first author on request.)

The second section gives a few details on the experimental data that were used to test the new program. The third section, the paper's main, describes the most important features of the new program. Future appendages to the program are briefly mentioned in the fourth and final section of this paper.

## II. THE TEST DATA

The test data came in the form of two *.hke files that contain crystallographic information which was extracted with the well-known electron crystallography program CRISP/ELD 2.1 [1] from an experimental electron diffraction spot pattern of a $Ba_3Nb_{16}O_{23}$ crystal [10,11]. (That electron diffraction pattern came with the CRISP/ELD program.) One of these files used the CRISP/ELD automatic default primitive indexing. The other *.hke file used an alternative rectangular-centered indexing that was selected manually. The original *.hke files were both restricted to a maximal spatial resolution of 1.25 Å, resulting in 256 spots.

The default primitive indexing mode of CRISP/ELD extracted the lattice parameters $a = 12.46 \pm 0.15$ Å, $b = 12.41 \pm 0.15$ Å, $\gamma = 119.5 \pm 1.0°$. The extracted lattice parameters for the alternative rectangular-centered indexing of the diffraction pattern came to $a = 12.45 \pm 0.15$ Å, $b = 21.60 \pm 0.15$ Å, and $\gamma = 89.5 \pm 1.0°$ (before symmetrization to 90°). Both lattices are oblique but within error bars also consistent with the lattice-metric symmetry restrictions of the rectangular-centered and hexagonal 2D Bravais lattices [15]. These two *.hke files allow for an internal consistency check of the calculations that the new program performed, because they contain essentially the same information (but slightly different spot intensities) and calculation results cannot depend on the labels of the electron diffraction spots.

## III. THE NEW COMPUTER PROGRAM

### A. The program's needed inputs and provided outputs

The new program only requires a single input: a plain text file with two columns of projected Laue indices (H and K) of electron diffraction spots and a column of extracted spot intensity values that correspond to each H,K index pair. There are numerous types of plain text files with crystallographic data such as *.hke, *.cif, etc. that contain unnecessary information in the form of additional columns (beyond of the three mentioned above) as well as additional information that, most often, is displayed as the top rows of the given text file. The program operates on all of these types of plain text files (as long as the input data is in the form of columns, alignment of the columns does not matter but order does) by giving the user the ability to remove unnecessary information. Thus, the user can simply tell the program which rows to delete as well as which columns are outside of the aforementioned three required columns.

The data in Table 1, a screenshot of one of the program's outputs, is based on the analysis of the *.hke file of the test data with the default primitive indexing by the CRISP/ELD program. The output starts with the name of the file that the user ran an analysis on. There is then an output that shows whether the program added HK index pairs with zero intensities for missing spots or not. In this particular table, 14 HK index pairs with spot intensities of zero were added which resulted in a total of 270 spot intensities.

The need to add index pairs is assessed by iterating through all of the Laue index pairs present in the original text file and checking if index pairs for all of the symmetries that are to be tested are present out to the maximal resolution of the electron diffraction spots given in the *.hke file. Another feature of the program is that it automatically creates HKi triples from Laue index pairs for calculations involving the symmetries of the hexagonal crystal system.

**Table 1**. Screenshot of results for the primitive (hexagonal) indexing. Calculation results are rounded from fifteen decimal places to four in the above table except for the $R_{sym}$ values which are rounded from fifteen decimal places to three (and then multiplied by 100).

```
File name:
256hex.txt
Number of spot intensities:
starting: 256  added: 14  total: 270
Best symmetry group:
2mm
```

| Site symmetry of geometric model | N-SSR | G-AIC value (bit) | Likelihood to be K-L best model | Akaike weight (%) | Rsym (%) |
|---|---|---|---|---|---|
| 2 | 0.8775 | 2.1429 | 0.7554 | 20.721 | 15.1 |
| ..m | 0.515 | 1.7805 | 0.9055 | 24.8382 | 11.5 |
| .m. | 0.5058 | 1.7713 | 0.9096 | 24.9525 | 11.9 |
| 2mm | 0.9491 | 1.5819 | 1.0 | 27.4313 | 16.0 |
| 3 | 9.3918 | 10.2355 | 0.0132 | 0.3624 | 56.0 |
| 3m1 | 9.4934 | 9.9153 | 0.0155 | 0.4253 | 56.0 |
| 31m | 9.5129 | 9.9347 | 0.0154 | 0.4212 | 56.0 |
| 6 | 9.6044 | 10.0262 | 0.0147 | 0.4023 | 56.1 |
| 6mm | 9.6094 | 9.8203 | 0.0163 | 0.4459 | 56.0 |

The last line of output before the table of information-theoretic metrics is the objectively determined oriented site/point symmetry group of the Kullback-Leibler (K-L) best geometric model of the digital input data, i.e. the most important symmetry classification measure. This symmetry group is determined using the hierarchy tree in Fig. 1, which results from the well-known symmetry inclusion relationships between maximal subgroups and minimal supergroups of the crystallographic 2D point groups. The inequality values shown in the hierarchy tree are determined using ratios of N-SSR values (see equation (1) below) as well as the number of symmetry operations a point symmetry group possesses, $k$, which is given on the left and right hand sides of this figure. There are two data tables in total (the first is given as Table 1) that list the new program's outputs.

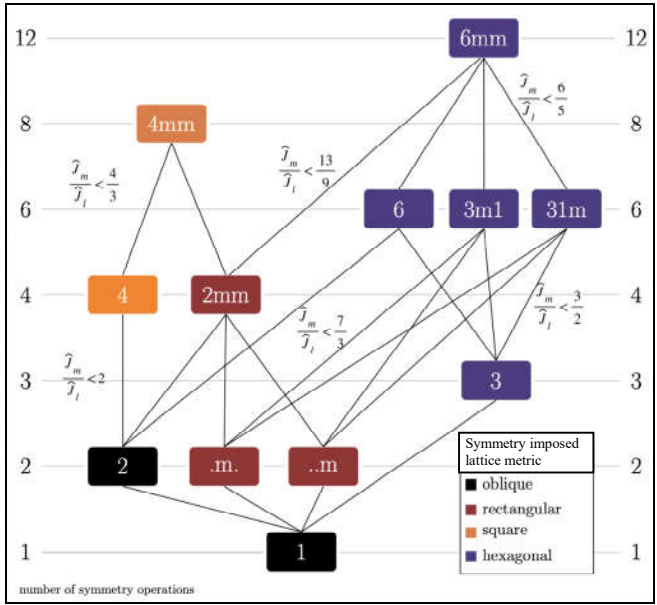

**Figure 1.** The hierarchy tree of oriented 2D site/point symmetries used to determine the geometric model that features the K-L best symmetry group.

The first table displays the N-SSR values, geometric Akaike Information Criteria (G-AIC) values in bits, the likelihood that a geometric model of the digital input data (with a certain point symmetry) is the K-L best model, the geometric Akaike weights in percent, and the traditional (crystallographic) $R_{sym}$ values [1] in percent. The second data table (not shown here) completes the program's output and gives the confidence levels (in percent) that one can move from a maximal subgroup to its minimal supergroup in the process of identifying the symmetry of the K-L best model.

### B. Equations and inequalities used in the production of Table 1

The equations, inequalities, and methods used for the identification of the K-L best geometric model (amongst all of the tested geometric models) for the experimental input data, columns two through five in Table 1, and the percent confidence calculations (result table not shown) will now be given. For an in depth explanation of the information-theoretic methodology for crystallographic symmetry classifications and quantifications in two dimensions, see [7-9].

The numerical pillar of the analysis by the new program are the normalized sums of squared residual (N-SSR) values of the competing geometric models of the input data. Normalization means in this context that before the SSR values are calculated, equation (1), for each of the nine geometric models that are to be tested, the spot intensities for each Laue index pair are divided by the largest spot intensity value in the input data file. (This is done to make numerical outputs easier to work with and does not affect the identification of the K-L best model.) After the program has appended missing index pairs with a spot intensity of zero, the "intensity SSR equation"

$$J = \sum_{j=1}^{N}(I_{j,\exp} - I_{j,sym})^2 \quad (1)$$

is used, with the $N$ spot intensities $J_j(H,K)$ from the input text file (after normalization) and $I_{sym}$ values as calculated according to [1]. They are used in conjunction with the inequalities shown in Figure 1 as obtained by

$$\frac{J_m}{J_l} < 1 + \frac{2(k_m - k_l)}{k_m(k_l - 1)} \quad (2),$$

with subscript $m$ representing the more symmetric/restricted model and subscript $l$ the less symmetric/restricted model. The variable $k$ stands for the number of symmetry operations in the point symmetry group, as given on the vertical axes of Fig. 1. (This inequality is the special $N_m = N_l$ version of a more general inequality in [7,9], where $N$ is the number of data points/index pairs.)

The program first identifies the "anchoring group" from which climbing up tests will proceed, using the symmetry inclusion relationships in Figure 1. This group is defined by the smallest N-SSR value amongst all of the model symmetries with $k = 2$ and 3. Then climbing up from this maximal subgroup to a minimal supergroup (and beyond) is permitted as long as inequality (2) is fulfilled. The process is repeated until no further climbing up is permitted by (2). For a minimal supergroup to be the symmetry of the K-L best model, inequality (2) must be fulfilled for *all* of its maximal subgroups, as well as for *all* of those subgroups' maximal subgroups etc. It is through the "climbing" up process from maximal subgroups to minimal supergroups, using the oriented site/point group hierarchy tree in Fig. 1, that the program determines the symmetry of the K-L best geometric model output. When inequality (2) is fulfilled the ad-hoc confidence level equation,

$$C_m = \frac{1 - K}{1 - K_{critical}} \cdot 100\% \quad (3),$$

is used which takes values between 100% and zero. Equations for $K$ and $K_{critical}$ are provided in [6,9]. A higher percentage represents a greater confidence that the N-SSR value ratios calculated by the program allow a climbing up from a given maximal subgroup to its minimal supergroup. For convenience, one uses the average confidence level for selecting a given minimal supergroup over all of its maximal subgroups.

First-order G-AIC values are model-selection bias-corrected symmetry quantifiers for rational, i.e. strictly data supported, model selections for small and modest amounts of generalized noise and calculated by

$$G - AIC = J + 2(dN + n)\hat{\varepsilon}^2 \quad (4)$$

where $J$ represents the N-SSR value for a given geometric model $S$, $d$ the dimension of $S$ (zero in our case), $N$ the number of data points, $n$ the number of degrees of freedom of $S$, and $\hat{\varepsilon}^2$ the variance of the generalized noise term. (The G-AIC values are listed in Table 1 in the adjacent column to the N-SSR values.) As expected, non-disjoint lower symmet-ric models always have lower N-SSR values due to greater degrees of freedom caused by fewer symmetry constraints. Kanatani's G-AIC accounts for this by a higher geometric model selection penalty term for lower symmetric models (with lower sums of squared residual). The variance of the generalized noise, which needs to be approximately Gaussian distributed for consistency, is calculated using

$$\widehat{\varepsilon}_{best}^2 \approx \frac{\widehat{J}_{best}}{rN_{best} - n_{best}} \quad (5),$$

in which *best* stands for the K-L best site/point symmetry model for the experimental input data and $r$ is the so-called co-dimension, which is in our case unity. Geometric Akaike Weights are calculated from

$$w_i = \frac{\exp(-\tfrac{1}{2}\Delta_i)}{\sum_{r=1}^{R} \exp(-\tfrac{1}{2}\Delta_r)} (100\%) \quad (6),$$

with

$$\Delta_i = G\text{-}AIC_i - G\text{-}AIC_{best} \quad (7).$$

The geometric Akaike weights are the probabilities that a certain geometric model is the K-L best model in the set of tested geometric models for the input data. The program's calculated Akaike weights add up to exactly 100%, as they should per definitions of Akaike weights and probabilities.

Columns five and six in Table 1 are for comparison purposes as high geometric Akaike weights can be said to "loosely correlate" with low values of traditional $R_{sym}$ calculations, which are computed as described in [1].

*C. Internal consistency check of the new program with the second *.hke file*

Table 2 gives the program's first output table for the *.hke file with the rectangular-centered indexing.

**Table 2.** Screenshot of the results for rectangular-centered indexing. Results are rounded in the same manner as in Table 1.

```
File name:
256rec.txt
Number of spot intensities:
starting: 256 added: 2 total: 258
Best symmetry group:
2mm
```

| Site symmetry of geometric model | N-SSR | G-AIC value (bit) | Likelihood to be K-L best model | Akaike weight (%) | Rsym (%) |
|---|---|---|---|---|---|
| 2 | 0.8736 | 2.1606 | 0.7589 | 20.3208 | 15.4 |
| ..m | 0.5229 | 1.81 | 0.9043 | 24.2149 | 11.8 |
| .m. | 0.534 | 1.8211 | 0.8993 | 24.0807 | 12.1 |
| 2mm | 0.9653 | 1.6088 | 1.0 | 26.7773 | 16.3 |
| 3 | 7.7665 | 8.6245 | 0.03 | 0.8023 | 52.0 |
| 3m1 | 7.8606 | 8.2896 | 0.0354 | 0.9485 | 52.1 |
| 31m | 7.8675 | 8.2965 | 0.0353 | 0.9453 | 52.1 |
| 6 | 7.9552 | 8.3842 | 0.0338 | 0.9047 | 52.2 |
| 6mm | 7.9584 | 8.1729 | 0.0376 | 1.0055 | 52.1 |

All numerical values in both tables are rather similar to each other. This is as they must, because essentially the same information has been processed in both cases (as the type of the indexing labels is immaterial). It is clear from both tables in this paper that the symmetry of the analyzed diffraction pattern is most likely *2mm* and that the crystal is highly unlikely to feature a hexagonal Bravais lattice [11].

IV. FUTURE DEVELOPMENTS OF THE PROGRAM

So far the new program is restricted to input data that are laid out on oblique, rectangular-centered, and hexagonal Bravais lattices. Future developments will be able to deal with the full hierarchy of the 2D Bravais lattice types [15].